\documentclass[runningheads,envcountsame]{llncs}

\usepackage{amsmath}
\usepackage{amssymb}
\usepackage{latexsym} 
\usepackage[all]{xypic}
\usepackage{listings}

\newcommand{\couple}[1]{\{#1\}}
\newcommand{\set}[1]{\{\,#1\,\}}
\newcommand{\card}[1]{|#1|}
\renewcommand{\deg}[2][]{\mathrm{deg}_{#1}(#2)}

\newcommand{\Ent}[1]{\mathcal{E}(#1)}

\newcommand{\mystar}[2]{\varsigma^{#1}_{#2}}

\newcommand{\glue}{Gl}

\newcommand{\derived}{\partial}

\newcommand{\lcollapse}[2][]{\,\mbox{$\overline{\bigoplus}$}^{#1}_{#2}}
\newcommand{\collapse}[2][]{\,\mbox{$\bigoplus$}^{#1}_{#2}}

\newcommand{\NLOGSPACE}{\mathrm{NLOGSPACE}}
\newcommand{\NP}{\mathrm{NP}}
\newcommand{\POL}{\mathrm{P}}
\newcommand{\EXP}{\mathrm{EXPTIME}}

\newcommand{\Lmu}{\mathbb{L}_{\mu}}
\newcommand{\K}{\mathbb{K}}

\newcommand{\ETWO}{{\cal U}_{2}}

\newcommand{\breath}{\vskip 6pt}

\title{Undirected Graphs of Entanglement 2}
\author{Walid Belkhir  \and  Luigi Santocanale}
\institute{%
Laboratoire d'Informatique Fondamentale de Marseille\\
Universit\'e de Provence
}

\begin{document}

\maketitle

\begin{abstract}
  Entanglement is a complexity measure of directed graphs that origins
  in fixed point theory. This measure has shown its use in designing
  efficient algorithms to verify logical properties of transition
  systems.  We are interested in the problem of deciding whether a
  graph has entanglement at most $k$. As this measure is defined by
  means of games, game theoretic ideas naturally lead to design
  polynomial algorithms that, for fixed $k$, decide the problem. Known
  characterizations of directed graphs of entanglement at most $1$
  lead, for $k = 1$, to design even faster algorithms.
  In this paper we give two distinct characterizations of
  \emph{undirected} graphs of entanglement at most $2$.  With these
  characterizations at hand, we present a linear time algorithm to
  decide whether an undirected graph has this property.
\end{abstract}

%%% Local Variables: 
%%% mode: latex
%%% TeX-master: "main"
%%% End: 

\section{Introduction}

% \newcommand{\anglais}[1]{\underline{#1}\footnote{Anglais}}
% \newcommand{\references}[1]{\underline{#1}\footnote{Add references}}
% \newcommand{\whocares}[1]{\{#1\}\footnote{Who cares ?}}
% \newcommand{\doyoumean}[2]{\underline{#1}\footnote{Do you mean
%     \emph{#2} ?}}
% \newcommand{\generic}[2]{\{#1\}\footnote{#2}}
% \newcommand{\myparagraph}[1]{\marginpar{#1.}}
% \newcommand{\refereesquestion}[1]{\marginpar{#1 ?}}
% \newcommand{\longsentence}[1]{\generic{#1}{Endless sentence}}

%% \myparagraph{What is entanglement} %%
Entanglement is a complexity measure of finite directed graphs
introduced in \cite{berwanger,BerwangerThesis} as a tool to analyze
the descriptive complexity of the Propositional Modal $\mu$-calculus.
Roughly speaking, its purpose is to quantify to what extent cycles are
intertwined in a directed graph. Its game theoretic definition -- by
means of robbers and cops -- makes it reasonable to consider
entanglement a generalization of the tree-width of undirected graphs
\cite{seymourthomas} to another kind of graphs, a role shared with
other complexity measures appeared in the literature
\cite{georg01,dirtreewidth,dtreewidth,dagtreewidth}.

%%\myparagraph{Context of entanglement. 
%%Why we do believe entanglementis worth studying} %%
A peculiar aspect of entanglement, and also our motivation for
studying it among the other measures, is its direct filiation from
fixed point theory. Its first occurrence takes place within the
investigation of the variable hierarchy
\cite{BerwangerGraLen02,BerwangerLen05} of the Propositional Modal
$\mu$-Calculus \cite{kozen}.  The latter, hereby noted $\Lmu$, is
nowadays a well known and appreciated logic, capable to express many
computational properties of transition systems while allowing their
verification in some feasible way. As a $\mu$-calculus \cite{AN}
$\Lmu$ increases the expressive power of Hennessy-Milner logic, i.e.
multimodal logic $\K$, by adding to it least and greatest fixed point
operators that bind monadic variables.  Showing that there are
$\mu$-formulas $\phi_{n}$ that are semantically equivalent to no
formula with less than $n$ bound variables is the variable hierarchy
problem for a $\mu$-calculus. Such a hierarchy is also meaningful in
the simpler setting of iteration theories \cite{bloomesik}.
% It has been observed that the variable
% hierarchy is orthogonal to the alternation depth hierarchy -- which
% crucially takes into account both the least and the greatest fixed
% point operators. As a matter of fact, the variable hierarchy can be
% defined and investigated as soon as just one fixed point operator is
% considered. Possibly, iteration theories \cite{bloomesik} are a more
% appropriate framework within which to study this hierarchy.

The relationship between entanglement and the number of bound
variables in a $\mu$-term might be too technical to be elucidated
here. Let us say, however, that entanglement roughly is a syntactic
analogous of the variable hierarchy, the latter being defined only
w.r.t. a given semantics. To argue in this direction, the relevant
fact is Proposition 14 of \cite{berwanger}, stating that the
entanglement of a directed graph is the minimal feedback of its finite
unravellings.

A second important topic in fixed point theory is the model checking
problem for $\Lmu$.  The main achievement of \cite{berwanger} states
that parity games whose underlying graphs have bounded entanglement
can be solved in polynomial time.  This is a relevant result for the
matter of verification, since model checking $\Lmu$ is reducible in
linear time to the problem of deciding the winner of a parity game.
% This result appears to be related to the variable hierarchy of
% $\Lmu$, even if we believe that the relationship still need to be
% clarified.  Nonetheless,
Berwanger's result calls for the problem of
\emph{deciding whether a graph has entanglement at most $k$}, a
problem which we address in this paper.  When settled, we can try to
exploit the main result of \cite{berwanger}, for example by designing
algorithms to model check $\Lmu$ that may perform well in practice. 
% An
% obvious procedure to solve parity games, which depends on having at
% disposal a fast algorithm to decide whether the entanglement of a
% parity game is small, is as follows.  If the entanglement of a parity
% game is small, then Berwanger's algorithm is used to solve it and,
% otherwise, a standard exponential algorithm \cite{zielonka,jurdzinski}
% is used. At present the complexity of the problem we address is
% unclear.  
We shall argue that, for fixed $k$, deciding whether a graph
has entanglement at most $k$ is a problem in the class $\POL$.  The
algorithms solving these problems can be combined to show that
deciding the entanglement of a graph is in the class $\EXP$. We have
no reasons to believe that the problem is in $\NP$. Let us mention on
the way that a problem that we indirectly address is that of solving
parity games on undirected graphs.  These games can be solved in
linear time if Eva's and Adam's moves alternate. Yet, the complexity
of the problem is not known if consecutive moves of the same player
are allowed. 
% The standard reduction to the alternating case -- which
% inserts opposite color positions in a non alternating edge -- does not
% work in the undirected setting.

% As a matter of fact we shall see that deciding, for
% fixed $k$, whether a graph has entanglement at most $k$ is
% polynomial. 
% % We believe that entanglement is the key concept that will bring
% % another insight to answer the open question on the exact complexity
% % of solving parity games and $\mu-$calculus model checking
% % \cite{Emerson93,obdrzalek03fast,Jur00:STACS,JPZ06}.
% We are interested in the problem of deciding if a graph has
% entanglement at most $k$.

%%\refereesquestion{What is done in this paper} 
In this paper we show that deciding whether an undirected graph $G$
belongs to $\ETWO$, the class of undirected graphs of entanglement at
most $2$, can be solved in time $O(\card{V_{G}})$.  We shall present
an algorithm that crucially depends on two characterizations of the
class $\ETWO$. One of them proceeds by forbidden subgraphs: an
undirected graph belongs to $\ETWO$ if and only if it does not contain
(i) a simple cycle of length strictly greater than $4$, (ii) a length
$3$ simple cycle whose vertices have all degree $3$, (iii) a length
$4$ simple cycle with two adjacent vertices of degree $3$. A second
characterization constructs the class $\ETWO$ from a class of atomic
graphs, called the \emph{molecules}, and an operation, the \emph{legal
  collapse}, that glues together two graphs along a prescribed pair of
vertices.
%%
% Given the family
% $\mathcal{F}_k$ of graphs of entanglement at most $k$, we ask : can
% one characterize graphs of $\mathcal{F}_k$ by excluding a set of
% subgraphs?. In this paper we give a positive answer for the family
% $\mathcal{F}_2$ of undirected graphs. We shall prove that
% $\mathcal{F}_2$ is completely characterized by excluding subgraphs of
% (i) simple cycles of length strictly greater than $4$, (ii) triangles
% of all vertices of degree $2$, and (iii) squares of two adjacent
% vertices of degree $2$. A natural question arises, can one give an
% explicit construction of the family $\mathcal{F}_k$? We have also a
% positive answer for the family $\mathcal{F}_2$ of undirected graphs.
% We shall give a complete algebraic construction of $\mathcal{F}_2$ by
% means of special graphs: \emph{molecules} and a simple algebraic
% operator : the \emph{collapse}, that is a kind of juxtaposition
% operator of graphs. Finally, with such a construction in hand we
% devise a linear time algorithm that decides whether an undirected
% graph has entanglement at most $2$.

The two characterizations may be appreciated on their own,
independently of the algorithm they give rise.  Entanglement is an
intrinsically dynamic concept, due to its game theoretic definition.
As such it is not an easy object of study, while the two
characterizations prepare it for future investigations with standard
mathematical tools. They also suggest that entanglement is a quite
robust notion, henceforth worth being studied independently of its
fix-point theoretic background.
As a matter of fact, some of the properties we shall encounter have
already been under focus: the combinatorial characterization exhibits
surprising analogies with the class of House-Hole-Domino free graphs,
see \cite{jamisonolariu,chepoi}, a sort of generalization of graphs
admitting a perfect elimination ordering.  These graphs arise as the
result of looking for wider notions of ordering for graphs that still
ensure nice computational properties.
On the other hand, the algebraic characterization recalls the well
known fact that graphs of fixed arbitrary tree-width may be
constructed by means of an algebra of pushouts and relabelings
\cite{courcelle2}.  The algebra of legal collapses suggests that, for
entanglement, it might be possible to develop an analogous generic
algebraic framework. It also points to standard graph theoretic ideas,
such as $n$-connectiveness, as the proper tools by which to analyze
entanglement.

% \refereesquestion{Why do you study undirected graphs 
%if what has to be
%   studied are directed graphs} 
Clearly, a work that still need to be carried out is to look for some
useful characterization of \emph{directed} graphs of entanglement at
most $k$.  At present, characterizations are known only for $k \leq 1$
\cite[Proposition 3]{berwanger}.  We believe that the results
presented here suggest useful directions to achieve this goal.  In
particular, a suggestive path is to generalize the algebra of
molecules and legal collapses to an undirected setting. This path
might be a feasible one considering that many scientists have recently
developed ideas and methods to lift some algebraic framework from an
undirected to a directed setting.  W.r.t. the algebra of entanglement,
a source of ideas might be the recent development of directed homotopy
theory from concurrency \cite{dihomotopy}.

\breath \textbf{Acknowledgement.} We would like to thank the anonymous
referees for their useful comments, and for suggesting how to obtain
the algorithm presented in Section \ref{sec:algo} out of the algebraic
framework introduced in Section \ref{sec:classzeta}.

%%% Local Variables: 
%%% mode: latex
%%% TeX-master: "main"
%%% End: 

%%\newpage
\section{Entanglement Games}

The entanglement of a finite digraph $G$, denoted $\Ent{G}$, was
defined in \cite{berwanger} by means of some games $\Ent{G,k}$, $k =
0,\ldots ,\card{V_{G}}$. The game $\Ent{G,k}$ is played on the graph
$G$ by Thief against Cops, a team of $k$ cops. The rules are as
follows. Initially all the cops are placed outside the graph, Thief
selects and occupies an initial vertex of $G$.  After Thief's move,
Cops may do nothing, may place a cop from outside the graph onto the
vertex currently occupied by Thief, may move a cop already on the
graph to the current vertex.  In turn Thief must choose an edge
outgoing from the current vertex whose target is not already occupied
by some cop and move there.  If no such edge exists, then Thief is
caught and Cops win.  Thief wins if he is never caught.  The
entanglement of $G$ is the least $k \in N$ such that $k$ cops have a
strategy to catch the thief on $G$. It will be useful to formalize
these notions.

\begin{definition}%[Entanglement game]
  The entanglement game $\Ent{G,k}$ of a digraph $G$ is defined by:
  \begin{itemize}
  \item Its positions are of the form $(v,C,P)$, where $v \in V_{G}$,
    $C \subseteq V_{G}$ and $\card{C} \leq k$, $P \in \{Cops,
    Thief\}$.
  \item Initially Thief chooses $v_{0} \in V$ and moves to
    $(v_0,\emptyset,Cops)$.
  \item Cops can move  from $(v,C,Cops)$ to $(v,C',Thief)$
    where $C'$ can be
    \begin{enumerate}
    \item $C$ : Cops skip,
    \item $C \cup\set{v}$ : Cops add a new Cop on the
      current position,
    \item $(C \setminus\set{x}) \cup \set{v}$ : Cops move a placed Cop
      to the current position.
    \end{enumerate}
  \item Thief can move from $(v,C,Thief)$ to $(v',C,Cops)$ if
    $(v,v') \in E_{G}$ and $v' \notin C$.
  \end{itemize}
  Every finite play is a win for Cops, and every infinite play is a win
  for Thief. We let
  \begin{align*}
    \Ent{G} & = \min \set{ k\,|\,\text{Cops have a winning strategy in  $\Ent{G,k}$}}\,.
  \end{align*}
%   The \emph{entanglement} of a digraph $G$, noted $\Ent{G}$, is the
%   minimum $k\geq 0$ such that Cops have a winning strategy in the
%   entanglement game $\Ent{G,k}$.
\end{definition}
It is not difficult to argue that there exist polynomial time
algorithms that, for fixed $k \geq 0$ decide on input $G$ whether
$\Ent{G} \leq k$. Such an algorithm constructs the game $\Ent{G,k}$
whose size is polynomial in $\card{V_{G}}$ and $\card{E_{G}}$, since
$k$ is fixed.  Since the game $\Ent{G,k}$ is clopen, i.e. it is a
parity game of depth $1$, it is well known \cite{jurdzinski} that such
game can be solved in linear time w.r.t. the size of the graph
underlying $\Ent{G,k}$.

In \cite{berwanger} the authors proved that $\Ent{G} = 0$ if and only
if it is $G$ is acyclic, and that $\Ent{G}\leq 1$ if and only if each
strongly connected component of $G$ has a vertex whose removal makes
the component acyclic. Using these results it was argued that deciding
whether a graph has entanglement at most $1$ is a problem in
$\NLOGSPACE$.
 
While wondering for a characterization of graphs of entanglement at
most $2$, we observed that such a question has a clear answer for
\emph{undirected} graphs. To deal with this kind of graphs, we recall
that an undirected edge $\couple{u,v}$ is just a pair $(u,v), \,(v,u)$
of directed edges. We can use the results of \cite{berwanger} to give
characterizations of undirected graphs of entanglement at most $1$. To
this goal, for $n \geq 0$ define the $n$-star of center $x_0$, noted
$\mystar{n}{x_0}$, to be the undirected graph $(V,E)$ where
$V=\set{x_0,a_1,...,a_n}$ and $E =\set{
  \couple{x_0,a_1},...,\couple{x_0,a_n}}$. More generally, say that a
graph is a star if it is isomorphic to some $\mystar{n}{x_0}$.  Then
we can easily deduce:
\begin{proposition}
  If $G$ is an undirected graph, then $\Ent{G} = 0$ if and only if
  $E_{G} = \emptyset$, and $\Ent{G} \leq 1$ if and only if $G$ is a
  disjoint union of stars.
\end{proposition}
 
To end this section we state a Lemma that later will be used often.
We remark that its scope does not restrict to undirected graphs.
\begin{lemma}
  \label{lemma:entsubgraph}
  If $H$ is a subgraph of $G$ then $\Ent{H} \le \Ent{G}$.
\end{lemma}
As a matter of fact, Thief can choose an initial vertex from $H$ and
then he can restrict his moves to edges of $H$. In this way he can
simulate a winning strategy from $\Ent{H,k}$ to a winning strategy in
$\Ent{G,k}$.

%%% Local Variables: 
%%% mode: latex
%%% TeX-master: "main"
%%% End: 

\section{Molecules, Collapses, and the Class $\zeta_2$}
\label{sec:classzeta}

In this section we introduce a class of graphs and prove that the
graphs in this class have entanglement at most 2. It will be the goal
of the next sections to prove that these are all the graphs of
entanglement at most 2.
\begin{definition}
  A \emph{molecule} $\theta^{\varepsilon , n}_{a,b}$, where
  $\varepsilon \in \set{0,1}$ and $n \ge 0$, is the undirected graph
  $(V,E)$ with $V=\set{a,b,c_1,...,c_n}$ and
  \begin{align*}
     E & = 
     \begin{cases}
       \set{\couple{a,c_1},...,\couple{a,c_n},
         \couple{b,c_1},...,\couple{b,c_n} }\,,  & 
       \varepsilon =0 ,\\
       \set{\couple{a,b},\couple{a,c_1},...,\couple{a,c_n},
         \couple{b,c_1},...,\couple{b,c_n} }\,,  & 
       \varepsilon =1 .
     \end{cases}
  \end{align*}
  The \emph{glue points} of a molecule $\theta^{\varepsilon ,
    n}_{a,b}$ are $a,b$. Its \emph{dead points} are $c_{1},\ldots
  ,c_{n}$.
\end{definition}

It is not difficult to prove that molecules have entanglement at most
$2$.
% \begin{align*}
%   \Ent{ \theta^{\varepsilon,n}_{a,b}} & = 
%   \begin{cases}
%     1, & \textrm{if $(\varepsilon=0 \textrm{ and }n=1)$ or $(\varepsilon=1 \textrm{ and } n=0)$} \\
%     2 & \textrm{otherwise}.
%   \end{cases}
% \end{align*}
\begin{definition}
  Let $G_1$ and $G_2$ be two undirected graphs with $V_{G_{1}} \cap
  V_{G_{2}}=\emptyset$, let $a_1 \in V_{G_{1}}$ and $a_2 \in
  V_{G_{2}}$. The \emph{collapse} of $G_1$ and $G_2$ on vertices $a_1$
  and $a_2$, denoted
  $
  G_1 \collapse[z]{a_1,a_2} G_2,
  $
  is the graph $G$ defined as follows:
  \begin{align*}
    V_{G} \;= \;& (V_{G_1}\setminus \set{a_1}) \cup
    (V_{G_2}\setminus\set{a_2})
    \cup\set{z}, \text{ where } z \not\in V_{G_{1}} \cup V_{G_{2}},\\
    E_{G} \;= \;& \set{ \couple{x_1,y_1}\in E_{G_{1}} \,|\, a_{1}
      \not\in\set{x_1,y_{1}} \,} \cup \set{ \couple{x_2,y_2}\in
      E_{G_{2}}\,|\, a_{2}
      \not\in\set{x_2,y_{2}}} \\
    & \cup \set{ \couple{ x,z }\,| \,\couple{ x,a_{1} } \in E_{G_1}
      \textrm{ or } \couple{ x,a_{2} } \in E_{G_2} }\,.
  \end{align*}
\end{definition}  
We remark that $\collapse{}$ is a coproduct in the category of pointed
undirected graphs and, for this reason, this operation is commutative
and associative up to isomorphism.  The graph $\eta$, whose
set of vertices is a singleton, is a neutral element.
% \begin{remark} 
%   Note also that $\Ent{G_1 \bigoplus G_2} \ge
%   max(\Ent{G_1}, \Ent{G_2})$, since $G_1$ (as well as
%   $G_2$) is a subgraph of $G_1 \bigoplus G_2$, up to isomorphism.
% \end{remark} 
As we have observed, a molecule is an undirected graph coming with a
distinguished set of vertices, its glue points. Let us call a pair
$(G,\glue)$ with $\glue \subseteq V_{G}$ a \emph{glue graph}.  For
glue graphs we can define what it means that a collapse is legal.
\begin{definition}
  If $G_{1},G_{2}$ are glue graphs, then we say that $G_1
  \collapse[z]{a,b} G_2$ is a \emph{legal} collapse if $a \in
  \glue_{G_1}$ and $ b \in \glue_{G_2}$. We shall then use the
  notation $G_1 \lcollapse[z]{a,b} G_2$ and define
  \begin{align*}
    \glue_{G_1 \lcollapse[z]{a,b} G_2} & =
    (\glue_{G_1}\setminus\set{a}) \cup (\glue_{G_2}\setminus\set{b})
    \cup \set{z}\,,
  \end{align*}
  so that $G_1 \lcollapse[z]{a,b} G_2$ is a glue graph.
\end{definition}
Observe that the graph $\eta$ can be made into  a unit for the
legal collapse by letting $\glue_{\eta} = V_{\eta}$. Even if the
operation $\lcollapse{}$ is well defined only after the choice of the
two glue points that are going to be collapsed, it should be clear
what it means that a family of glue graphs is closed under legal
collapses.
\begin{definition}
  We let $\zeta_2$ be the least class of glue graphs containing the
  molecules, the unit $\eta$, and closed under legal collapses and
  graph isomorphisms.   
\end{definition} 
We need to make precise some notation and terminology. Firstly we
shall abuse of notation and write 
\begin{align*}
  G & = H \lcollapse{v} K
\end{align*}
to mean that there exist subgraphs $H,K$ of $G$ such that $v \in
\glue_{G} \cap V_{H} \cap V_{K}$ and $G$ is isomorphic to the legal
collapse $H \lcollapse[z]{v,v} K$. Notice that if $H$ and $K$ are
distinct from $\eta$, then $v$ is an articulation point of $G$.
Second, we shall say that a graph $G$
belongs to $\zeta_{2}$ to mean that there exists a subset $\glue
\subseteq V_{G}$ such that the glue graph $(G,\glue)$ belongs to
$\zeta_{2}$. We can now state the main result of this section.
\begin{proposition}%
  \label{zeta2algif} 
  If $G$ belongs to the class $\zeta_2$, then $\Ent{G} \leq 2$.
\end{proposition}
\begin{proof}
  Observe that, given a molecule $\theta^{\varepsilon,n}_{a,b}$
  occurring in an algebraic expression for $G$, we can rearrange the
  summands of the algebraic expression to write
  \begin{align}
    \label{eq:molleftright}
    G & = L \lcollapse{a} \theta^{\varepsilon,n}_{a,b}
    \lcollapse{b} R 
  \end{align}
  where $L,R \in \zeta_{2}$.  A Cops winning strategy in the game
  $\Ent{G,2}$ is summarized as follows. If Thief occupies some vertex
  of the molecule $\theta^{\varepsilon,n}_{a,b}$, Cops will place its
  two cops on $a$ and $b$, in some order.  By doing that, Cops will
  force Thief to move (i) on the left component $L$, in which case
  Cops can reuse the cop on $b$ on $L$, (ii) on the molecule
  $\theta^{\varepsilon,n}_{a,b}$, in which case Thief will be caught
  in a dead point of the molecule, (iii) on the right component $R$,
  in which case Cops can reuse the cop on $a$ on $R$.
  
  Cops can recursively use the same strategy in $\Ent{L,2}$ and
  $\Ent{R,2}$. The recursion terminates as soon as in the expression
  \eqref{eq:molleftright} for $G$ we have $L = R = \eta$.
  \qed
\end{proof}

The reader will have noticed similarities between the strategy
proposed here and the strategy needed in \cite{berwanger} to argue
that undirected trees have entanglement at most $2$. As a matter of
fact, graphs in $\zeta_{2}$ have an underlying tree structure. For a
glue graph $G$, define the derived graph $\derived G$ as follows: its
vertices are the glue points of $G$, and $\couple{a,b} \in E_{\derived
  G}$ if either $\couple{a,b} \in E_{G}$ or there exists $x \in
V_{G}\setminus \glue_{G}$ such that $\couple{a,x},\couple{x,b} \in
E_{G}$. The following Proposition is not difficult to prove.
\begin{proposition}
  \label{prop:derived}
  A glue graph $G$ is in $\zeta_{2}$ if and only if $\derived G$ is a
  forest, and each $x \in V_{G} \setminus \glue_{G}$ has exactly two
  neighbors, which moreover are glue points.
\end{proposition}

%%% Local Variables: 
%%% mode: latex
%%% TeX-master: "main"
%%% End: 

\section{Combinatorial Properties}
The goal of this section is to setup the tools for the
characterization Theorem \ref{theo:zeta2alg}.  We deduce some
combinatorial properties of undirected graphs of entanglement at most
$2$. To this goal, let us say that a simple cycle is long it its
length is strictly greater than $4$, and say otherwise that it is
short.  Also, let us call a simple cycle of length $3$ (resp.  $4$) a
triangle (resp. square).
\begin{proposition}%
  \label{zeta2com}
  \label{prop:zeta2com}
  An undirected graph $G$ such that $\Ent{G}\leq 2$ satisfies the
  following conditions:
  \begin{align*}
    -\;& \text{a simple Cycle of $G$ is Short,} \label{cond:shortcycles}\tag{CS} \\
    -\;& \text{a triangle of $G$ has at least one vertex of
      degree $2$,}
    \label{cond:no3colls}
    \tag{No-3C} \\
    -\;& \text{a square of $G$ cannot have two adjacent vertices} \\
    &\mbox{\hspace{50mm}}\text{% 
        of degree strictly greater than $2$.}
    \label{cond:noadjcolls}
    \tag{No-AC} 
  \end{align*}
\end{proposition}
Condition \eqref{cond:no3colls} forbids as subgraphs of $G$ the graphs
arising from the scheme on the left of figure 1. These are made up of
a triangle and 3 distinct Collapses, with vertices $x,y,z$ that might
not be distinct.  Condition \eqref{cond:noadjcolls} forbids the scheme
on the right of figure 1, made up of a square and two Adjacent
Collapses, with vertices $x,y$ that might not be distinct.
Let us remark that graphs satisfying \eqref{cond:shortcycles},
\eqref{cond:no3colls}, and \eqref{cond:noadjcolls} are
House-Hole-Domino free, in the sense of \cite{chepoi}. With respect to
$HDD$-free graphs, the requirement is here stronger since for example
long cycles are forbidden as subgraphs, not just as induced subgraphs.
\begin{figure}
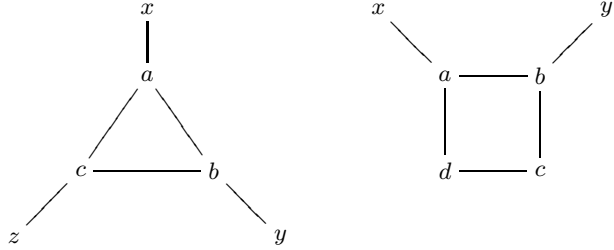

  $$
  \xygraph{ []*+{a}="a" (%
    -[r(0.7)d]*+{b}="b", -[l(0.7)d]*+{c}="c" )%
    "b"-"c" "a"-[u(0.7)]*+{x} "b"-[r(0.7)d(0.7)]*+{y}
    "c"-[l(0.7)d(0.7)]*+{z} }%
  \;\;\;\;\;\; %
  \xygraph{ %
    []*+{a}="a"
    -[r]*+{b}="b" -[d]*+{c}="c" -[l]*+{d}="d" -"a"
    "a"-[l(0.7)u(0.7)]*+{x} "b"-[r(0.7)u(0.7)]*+{y} }
  $$\centering
  \label{figure:3CAC}
  \caption{The graphs $3C$ and $AC$}
\end{figure}  
We shall see with Theorem \ref{zeta2alg} that these properties
completely characterize the class of undirected graphs of entanglement
at most $2$.  Proposition \ref{prop:zeta2com} is an immediate
consequence of Lemma \ref{lemma:entsubgraph} and of the following
Lemmas \ref{lemma:cycles}, \ref{lemma:3C}, \ref{lemma:AC}.

Let $P_{0}$ be the empty graph and, for $n \geq 1$, let $P_{n}$ be the
path with $n$ vertices and $n-1$ edges: $V_{P_{n}}=\set{0,...,n-1}$
and $\couple{i,j} \in E_{P_{n}}$ iff $|i-j| = 1$.  For $n \geq 3$, let
$C_n$ be the cycle with $n$ vertices and edges:
$V_{C_{n}}=\set{0,...,n-1}$ and $\couple{i,j} \in E_{C_{n}}$ iff
$|i-j| \equiv 1 \mod n$.
\begin{lemma}
  \label{lemma:cycles}
  If $n \geq 5$ then $\mathcal{E}(C_n) \geq 3$. 
\end{lemma}  
\begin{proof}  
  To describe a winning strategy for Thief in the game $\Ent{C_n,2}$
  consider that the removal of one or two vertices from $C_{n}$
  transforms such graph into a disjoint union $P_{i} + P_{j}$ with $i +
  j \geq n - 2 \geq 3$: notice in particular that $i \geq 2$ or $j
  \geq 2$.  In a position of the form $(v,C,Thief)$ with $v \in C$,
  Thief moves to a component $P_{i}$ with $i \geq 2$.  From a position
  of the form $(v,C,Thief)$ with $v \not\in C$, $v$ in some
  component $P_{i}$, and $i \geq 2$, Thief moves to some other vertex
  in the same component.  This strategy can be iterated infinitely
  often, showing that Thief will never be caught.  \qed
\end{proof} 

\begin{lemma}
  \label{lemma:3C} 
  Let $3C$ be a graph on the left of figure 1.
  % \ref{figure:3CAC}.
  We have $\Ent{3C} \geq 3$.
\end{lemma}
\begin{proof}
  A winning strategy for Thief in the game $\Ent{3C,2}$ is as follows.
  By moving on $a,b,c$, Thief can force Cops to put two cops there,
  say for example on $a$ and $b$. Thief can then escape to $c$ and
  iterate moves on the edge $\couple{c,z}$ to force Cops to move one
  cop on one end of this edge. From a position of the form
  $(c,C,Thief)$ with $c \in C$, Thief moves to a free vertex among
  $a,b$.  From a position of the form $(z,C,Thief)$ with $c \not\in C$
  Thief moves to $c$ and forces again Cops to occupy two vertices
  among $a,b,c$. Up to a renaming of vertices, such a strategy can be
  iterated infinitely often, showing that Thief will never be caught.

  Observe that the proof does not depend on $x,y,z$ being distinct.
  \qed
\end{proof}

\begin{lemma}
  \label{lemma:AC}
  Let $AC$ be a graph on the right of figure
  1. %\ref{figure:3CAC}.
  We have $\Ent{AC} \geq 3$.
\end{lemma}
\begin{proof}
  By moving on $a,b,c,d$, Thief can force Cops to put two cops either
  on $a,c$ or on $b,d$: let us say $a,c$. Thief can then escape to $b$
  and iterate moves on the edge $\couple{b,y}$ to force Cops to move
  one cop on one end of this edge. From a position of the form
  $(b,C,Thief)$ with $b \in C$, Thief moves to a free vertex among
  $a,c$.  From a position of the form $(y,C,Thief)$ with $b \not\in C$
  Thief moves to $b$ and forces again Cops to occupy either $a,c$ or
  $b,d$. Up to a renaming of vertices, such a strategy can be iterated
  infinitely often, showing that Thief will never be caught.
  Again, we observe that the strategy does not depend on $x,y$ being
  distinct.  \qed
\end{proof}  
We end this section by pointing out that $\Ent{C_{n}} = \Ent{3C} =
\Ent{AC} = 3$ ($n \geq 5$).

%%% Local Variables: 
%%% mode: latex
%%% TeX-master: "main"
%%% End: 

\section{Characterization of Entanglement at Most 2}
\label{subsec:converse}

In this section we accomplish the characterization of the class of
undirected graphs of entanglement at most $2$: we prove that this
class coincides with $\zeta_2$.

The following Lemma is the key observation by which the induction
works in the proof of Proposition \ref{zeta2algonly}.  It is worth,
before stating it, to recall the difference between $\collapse{}$, the
collapse of two ordinary undirected graphs, and
$\lcollapse{}$, the legal collapse of two glue graphs.
\begin{lemma} 
  \label{lemma:induct1} Let $G$ be an undirected graph satisfying
  \eqref{cond:no3colls} and \eqref{cond:noadjcolls}.  If $G =
  \theta_{v,b}^{\varepsilon,n} \collapse{b} H$ and $H \in \zeta_2$,
  then there is a subset $\glue' \subseteq V_{G}$ such that
  $(H,\glue')$ is a glue graph in $\zeta_{2}$, $b \in \glue'$, and
  moreover $G$ is the result of the legal collapse 
  $G  = \theta_{v,b}^{\varepsilon,n} \lcollapse{b} (H,\glue')$.
%   \begin{align*}
%     G & = \theta_{v,b}^{\varepsilon,n} \lcollapse{b}
%     (H,\glue')\,. 
%   \end{align*}
  Consequently, $G \in \zeta_2$, with $v$ a glue point of $G$.
\end{lemma}
The proof of the Lemma doesn't present difficulties and therefore it
omitted.
% it
% is delayed to the appendix.

\begin{proposition} %%
  \label{zeta2algonly} %%
  If $G$ is an undirected graph satisfying \eqref{cond:shortcycles},
  \eqref{cond:no3colls}, and \eqref{cond:noadjcolls}, then $G \in
  \zeta_2$.
\end{proposition}
\begin{proof}
  The proof is by induction on $\card{V_{G}}$. Clearly the Proposition
  holds if $\card{V_{G}} =1$, in which case $G = \eta \in \zeta_{2}$.
  Let us suppose the Proposition holds for all graphs $H$ such that
  $\card{V_{H}} < \card{V_{G}}$.

  If all the vertices in $G$ have degree less than or equal to $2$,
  then $G$ is a disjoint union of paths and cycles of length at most
  $4$. Clearly such a graph belongs to $\zeta_{2}$.  Otherwise, let
  $v_0$ be a vertex such that $\deg[G]{v_{0}} \geq 3$ and consider the
  connected components $G_{\ell}$, $\ell = 1,\ldots ,h$, of the graph
  $G \setminus \set{v_{0}}$.  Let $G_{\ell}^{v_{0}}$ be the subgraph
  of $G$ induced by $V_{G_{\ell}} \cup \set{v_0}$. We shall show that
  this graph is of the form
  \begin{align}
    \label{eq:decoupage}
    G_{\ell}^{v_{0}} & = \theta^{\varepsilon,m}_{v_{0},v_{1}}
    \collapse{v_{1}} H\,,
  \end{align}
  for some $\varepsilon \in \set{0,1}$, $m \geq 0$, and a graph $H \in
  \zeta_{2}$.

  Clearly, if $G_{\ell}$ is already a connected component of $G$, then
  $G_{\ell} \in \zeta_{2}$ by the inductive hypothesis.  We can pick
  any $v_{1} \in V_{G_{\ell}}$ and argue that formula
  \eqref{eq:decoupage} holds with $m = \varepsilon = 0$, $H =
  G_{\ell}$.

  Otherwise, let $\mathcal{N}_{\ell}=\set{a_1,...,a_n}$, $n \geq 1$,
  be the set of vertices of $G_{\ell}^{v_{0}}$ at distance $1$ from
  $v_0$. We claim that either the subgraph of $G_{\ell}$ induced by
  $\mathcal{N}_{\ell}$, noted $\mathcal{N}_{G_{\ell}}$, is a star or
  there exists a unique $v_1 \in G_{\ell}$ at distance $1$ from
  $\mathcal{N}_{\ell}$, and moreover the subgraph of $G_{\ell}$
  induced by $\mathcal{N}_{\ell}\cup \set{v_1}$ is a star. In both
  cases, a vertex of such a star which is not the center has degree
  $2$ in $G$.

  (i) If $E_{\mathcal{N}_{G_{\ell}}} \neq \emptyset$, then
  $\mathcal{N}_{G_{\ell}}$ is a star. Let us suppose that
  $\couple{a_{1},a_{2}} \in E_{G_{\ell}}$.  Since $G_{\ell}$ is
  connected, if $a_k \in \mathcal{N}_{\ell} \setminus \set{a_1,a_2}$
  then there exists a path from $a_{k}$ to both $a_1$ and $a_2$.
  Condition \eqref{cond:shortcycles} implies that either
  $\couple{a_{1},a_{k}} \in E_{G_{\ell}}$, or $\couple{a_{k},a_{2}}
  \in E_{G_{\ell}}$. If $x_0 \in V_{G_{\ell}} \setminus \set{a_{2}}$
  then there cannot be a simple path $a_k \ldots x_0 \ldots a_1$
  otherwise $v_0a_k\ldots x_0\ldots a_1 a_2 v_0$ is a long cycle.
  Therefore, a simple path from $a_{k}$ to $a_{1}$ is of the form $a_k
  a_1$ or $a_k a_2 a_1$. By condition \eqref{cond:no3colls} it is not
  the case that $\couple{a_{k},a_{1}},\couple{a_{k},a_{2}} \in
  E_{G_{\ell}}$, otherwise $\set{v_{0},a_{1},a_{2},a_{k}}$ is a clique
  of cardinality $4$.  Finally, if $\couple{a_{k},a_{1}} \in
  E_{G_{\ell}}$ and $a_{l} \in \mathcal{N}_{\ell} \setminus
  \set{a_1,a_2,a_{k}}$, then $\couple{a_{l},a_{1}} \in E_{G_{\ell}}$
  as well, by condition \eqref{cond:shortcycles}, otherwise
  $v_{0}a_{k}a_{1}a_{2}a_{l}v_{0}$ is a long cycle.  Therefore, if
  $\card{\mathcal{N}_{\ell}} > 2$, then $\mathcal{N}_{G_{\ell}}$ is a
  star with a prescribed center, which we can assume to be $a_{1}$.
  Since $\deg[G]{v_{0}} \geq 3$, by condition \eqref{cond:no3colls}
  only $a_{1}$ among vertices in $\mathcal{N}_{\ell}$ may have degree
  greater than $2$.  Otherwise $\card{\mathcal{N}_{\ell}} = 2$ and
  again at most one among $a_{i}$, $i =1,2$, has $\deg[G]{a_{i}} > 2$.
  Again, we can assume that $\deg[G]{a_{2}} = 2$.  We deduce that the
  subgraph of $G^{v_{0}}_{\ell}$ induced by $\set{v_{0}} \cup
  \mathcal{N}_{\ell}$ is of the form $\theta^{1,n-1}_{v_{0},a_{1}}$.

  (ii) If $E_{\mathcal{N}_{G_{\ell}}}= \emptyset$, then we distinguish
  two cases. If $\card{\mathcal{N}_{\ell}}=1$, then the subgraph of
  $G^{v_{0}}_{\ell}$ induced by $\set{v_{0}} \cup \mathcal{N}_{\ell}$
  is $\theta^{1,0}_{v_{0},a_{1}}$. Otherwise, if
  $\card{\mathcal{N}_{\ell}} \ge 2$, between any two distinct vertices
  in $\mathcal{N}_{\ell}$ there must exist a path in $G_{\ell}$, since
  $G_{\ell}$ is connected.  By condition \eqref{cond:shortcycles}, if
  $a_{i}\ldots x_{i,j}\ldots a_{j}$ is a simple path from $a_{i}$ to
  $a_{j}$ with $x_{i,j} \in V_{G_{\ell}}\setminus \mathcal{N}_{\ell}$,
  then $\couple{a_{i},x_{i,j}},\couple{a_{j},x_{i,j}} \in
  E_{G_{\ell}}$.  Also \eqref{cond:shortcycles} implies that, for
  fixed $i$, $x_{i,k} = x_{i,j}$ if $k \neq j$, otherwise
  $v_{0}a_{k}x_{i,k}a_{i}x_{i,j}a_{j}v_{0}$ is a long cycle.  We can
  also assume that $x_{i,j} = x_{j,i}$, and therefore $x_{i,j} =
  x_{i,k} = x_{l,k}$ whenever $i \neq j$ and $l \neq k$.  Thus we can
  write $x_{i,j} = v_{1}$ for a unique $v_{1}$ at distance $2$ from
  $v_{0}$.  Since $\card{\mathcal{N}_{\ell}} \ge 2$ and
  $\deg[G]{v_{0}} \geq 3$, condition \eqref{cond:noadjcolls} implies
  that $\deg[G]{a_{i}} = 2$ for $i = 1,\ldots ,n$. We have shown that
  in this case the subgraph of $G^{v_{0}}_{\ell}$ induced by
  $\mathcal{N}_{\ell} \cup \set{v_{0},v_{1}}$ is a molecule
  $\theta^{0,n}_{v_{0},v_{1}}$, with $n \geq 2$.

  Until now we have shown that \eqref{eq:decoupage} holds with $H$ a
  graph of entanglement at most $2$. Since for such a graph
  $\card{V_{H}} < \card{V_{G}}$, the induction hypothesis implies $H
  \in \zeta_{2}$.  Lemma \ref{lemma:induct1} in turn implies that
  $G_{\ell}^{v_{0}} \in \zeta_{2}$, with $v_{0}$ a glue point of
  $G_{\ell}^{v_{0}}$. Finally we can use
  \begin{align*}
    G & = G^{v_{0}}_1 \lcollapse{v_{0}} G^{v_{0}}_2
    \lcollapse{v_{0}} ...  \lcollapse{v_{0}}
    G^{v_{0}}_h\,,
  \end{align*}
  to deduce that $G \in \zeta_{2}$.
  \qed
\end{proof} 

We can now state our main achievement.
\begin{theorem}
  \label{zeta2alg}
  \label{theo:zeta2alg}
  For a finite undirected graph $G$, the following are equivalent:
  \begin{enumerate}
  \item $G$ has entanglement at most $2$,
  \item $G$ satisfies conditions \eqref{cond:shortcycles},
    \eqref{cond:no3colls}, \eqref{cond:noadjcolls},
  \item $G$ belongs to the class $\zeta_2$.
  \end{enumerate} 
\end{theorem}
As a matter of fact, we have shown in the previous section that 1
implies 2, in this section that 2 implies 3, and in section
\ref{sec:classzeta} that 3 implies 1.

% \breath

% To end this section, we add a remark on the tree width of undirected
% graphs in the class $\zeta_{2}$. The reader may have noticed that
% graphs in $\zeta_{2}$ have tree width at most $2$. However, in view of
% \cite[Proposition 8]{berwanger}, not every graph of tree width $2$ has
% entanglement $2$: in particular the tree width of the graph 3C is $2$,
% its entanglement is $3$.

 %%% Local Variables: 
 %%% mode: latex
 %%% TeX-master: "main"
 %%% End:  

\section{A Linear Time  Algorithm}
\label{sec:algo}
In this section we present a linear time algorithm that decides
whether an undirected graph $G$ has entanglement at most $2$.  We
would like to thank the anonymous referee for pointing to us the ideas
and tools needed to transform the algebraic characterization of
Section \ref{sec:classzeta} into a linear time algorithm.

Let us recall that, for $G = (V,E)$ and $v \in V$, $v$ is an
\emph{articulation point} of $G$ iff there exist distinct $v_{0},v_{1}
\in V \setminus \set{v}$ such that every path from $v_{0}$ to $v_{1}$
visits $v$. Equivalently, $v$ is an \emph{articulation point} iff the
subgraph of $G$ induced by $V \setminus \set{v}$ is disconnected.
The graph $G$ is \emph{biconnected} if it does not contain
articulation points. A subset of vertices $V' \subseteq V$ is
biconnected iff the subgraph induced by $V'$ is biconnected. A
\emph{biconnected component} of $G$ is biconnected subset $C \subseteq
V$ such that if $C \subseteq V'$ and $V'$ is biconnected then $C = V$.
The \emph{superstructure} of $G$ is the graph $F_{G}$ defined as
follows. Its set of vertices is the disjoint union $V_{F_{G}} = {\cal
  A}(G) \uplus {\cal C}(G)$, where
\begin{align*}
  {\cal A}(G) & = \set{a \in V \mid a \text{ is an
      articulation point of G} }\,, \\
  {\cal C}(G) & = \set{C
    \subseteq V\mid C \text{ is a biconnected component of } G}\,, &
  \intertext{
    and its set of edges is of the form} E_{F_{G}} & =
  \set{\couple{a,C} \mid a \in {\cal A}(G), C \in {\cal C}(G), \text{
      and } a \in C}\,.
\end{align*}
It is well known that $F_{G}$ is a forest and that Depth-First-Search
techniques may be used to compute the superstructure $F_{G}$ in time
$O(\card{V} + \card{E})$, see \cite[\S 23-2]{Algorithms}. Observe also
that this implies that $\sum_{C \in {\cal C}(G)} \card{C}  = O(\card{V} + \card{E})$.
% \begin{align*}
%   \sum_{C \in {\cal C}(G)} \card{C} & = O(\card{V} + \card{E})\,.
% \end{align*}
This relation that may also be derived considering that biconnecetd
components do not share common edges, so that $\card{V_{F_{G}}} =
O(\card{V} + \card{E})$ and $\card{E_{F_{G}}} = O(\card{V} +
\card{E})$ since $F_{G}$ is a forest. We have therefore
\begin{align*}
  \sum_{C \in {\cal C}(G)} \card{C} & =
  \card{V\setminus {\cal A}(G)} + \sum_{a \in {\cal A}(G)}
  \card{\set{C \in {\cal C}(G)\mid a \in C}} \\
  & =
  \card{V\setminus {\cal A}(G)} + 
  \card{E_{F_{G}}} =
  O(\card{V} + \card{E})\,.
\end{align*}

The algorithm ENTANGLEMENT-TWO relies on the following considerations.
If a graph $G$ belongs to the class $\zeta_{2}$, then it has an
algebraic expression explaining how to construct it using molecules as
building blocks and legal collapses as operations. We can assume that
in this expression the molecule $\theta^{0,1}_{a,b}$ does not appear,
since each such occurrence may be replaced by the collapse
$\theta^{1,0}_{a,x} \lcollapse{x} \theta^{1,0}_{x,b}$. W.r.t.  this
normalized expression, if $G$ is connected then its articulation
points are exactly those glue points $v$ of $G$ that appears in the
algebraic expression as subscripts of some legal collapse
$\lcollapse{v}$; the molecules are the biconnected components of $G$.

The algorithm computes the articulation points and the biconnected
components of $G$ -- that is, its superstructure -- and afterwards it
checks that each biconnected component together with its articulation
points is a molecule.
\lstset{
  morekeywords={Procedure,Function,%
    let, if,then,else,not,begin,end,for,foreach,do,done,and,%
    otherwise, %%
    %%Input,
    return,
    accept, reject,
    true,false,
  },
  morecomment=[l][itshape]{//},
  mathescape=true,
  xleftmargin=8mm,
  %%frame=single,
  %%framexleftmargin=0mm
  numbers=left
}

\small
\begin{lstlisting}[]
ENTANGLEMENT-TWO($G$) 
// Input an undirected graph $G$, accept if $G \in \zeta_2$
if $\card{E} \geq  3\card{V}$ then reject
foreach $v \in V$ do $\deg{v} := \card{vE}$
let  $F_{G} = ({\cal A}(G) \uplus {\cal G}(G),E_{F_{G}})$ be the superstructure of $G$
foreach $C \in {\cal A}(G)$ do 
  if not IS-MOLECULE($C,\set{a \in {\cal A}(G) \mid a \in C}$) then reject
accept           
\end{lstlisting}
\normalsize %

For a biconnected component together with a set of candidate glue
points to be a molecule we need of course these candidates to be at
most 2. Also, every vertex whose degree in $G$ is not $2$ is a
candidate glue point. Improving on these observations we arrive at the
following characterization.
\begin{lemma}
  Let $G = (V,E)$ be a biconnected graph and $D \subseteq V$ be such
  that $\set{v \in V \mid \deg{v} \neq 2} \subseteq D$.
  Then $G$ is isomorphic to a molecule $\theta^{\epsilon,n}_{a,b}$,
  with $D$ isomorphically sent to a subset of $\set{a,b}$, if and only
  if either (i) $\card{D} = 2$ and $\couple{x,d} \in E$ for each $x
  \in V \setminus D$ and $d \in D$ or (ii) $\card{D} < 2$ and
  $\card{V} \in \set{3,4}$.
\end{lemma}
Therefore the recognition algorithm for a molecule is as follows.
\small
\begin{lstlisting}
IS-MOLECULE($C,A$)
if $\card{A} > 2$ then return false
let $D = \set{x \in C \mid \deg{x} \neq 2} \cup A$
if $\card{D} > 2$ then return false
if $\card{D} < 2$ then
  if $\card{C} \in \set{3,4}$ then return true
  else return false
foreach $x \in C \setminus D$ 
  if $D \not\subseteq xE$ then return false
return true   
\end{lstlisting}
\normalsize %

Let us now argue about time resources of this algorithm.
\spnewtheorem*{fact}{Fact}{\bfseries}{\itshape}
\begin{fact}
  Algorithm ENTANGLEMENT-TWO($G$) runs in time $O(\card{V_{G}})$.
\end{fact}
It is clear that the function IS-MOLECULE runs in time $O(\card{C})$,
so that the loop (lines 7-8) of ENTANGLEMENT-TWO runs in time
$O(\sum_{C \in {\cal C}(G)}\card{C}) = O(\card{V} + \card{E})$.
Therefore the algorithm requires time $O(\card{V} + \card{E})$.

The following Lemma, whose proof depends on considering a tree with
back edges arising from a Depth-First-Search on the graph, elucidates
the role of the 3rd line of the algorithm.
\begin{lemma}
  If a graph $(V,E)$ does not contain a simple cycle $C_{n}$ with $n
  \geq k$, then it has at most $(k-2)\card{V}-1$ undirected edges.
\end{lemma}
Line 3 ensures $\card{E_{G}} = O(\card{V_{G}})$ and that the algorithm
runs in time $O(\card{V_{G}})$.

 %%% Local Variables: 
 %%% mode: latex
 %%% TeX-master: "main"
 %%% End:  

\newpage

\bibliographystyle{splncs}
\bibliography{biblio}

% \newpage
% \appendix
% \input appendix.tex

\end{document}